%% file: LED_PRD_new.tex
\RequirePackage{lineno}

\documentclass[aps,prl,twocolumn,showpacs,superscriptaddress,groupedaddress]{revtex4}

\usepackage{graphicx}
\usepackage{dcolumn}
\usepackage{bm}
\usepackage{amssymb}
\usepackage{amsmath}
\usepackage{upgreek}
\usepackage{color}
\usepackage{multirow}
\usepackage{tikzpagenodes}
\usepackage[para]{threeparttablex}
\usepackage{fancyhdr}

\begin{document}

\widetext

\title{
\thispagestyle{fancy} 
\fancyhead{}
\fancyfoot{}
\fancyhead[R]{FERMILAB-PUB-16-172-ND}
Constraints on Large Extra Dimensions from the MINOS Experiment
}
\input minos-plus_feb2016.tex
\date{\today}

\begin{abstract}
We report new constraints on the size of large extra dimensions from data collected by the MINOS experiment between 2005 and 2012. Our analysis employs a model in which sterile neutrinos arise as Kaluza-Klein states in large extra dimensions and thus modify the neutrino oscillation probabilities due to mixing between active and sterile neutrino states. Using Fermilab's NuMI beam exposure of $10.56 \times 10^{20}$ protons-on-target, we combine muon neutrino charged current and neutral current data sets from the Near and Far Detectors and observe no evidence for deviations from standard three-flavor neutrino oscillations. The ratios of reconstructed energy spectra in the two detectors constrain the size of large extra dimensions to be smaller than $0.45\,\upmu\text{m}$ at 90\%\,C.L. in the limit of a vanishing lightest active neutrino mass. Stronger limits are obtained for non-vanishing masses.
\end{abstract}

\maketitle

Neutrino oscillation has been established through measurements of solar, atmospheric, reactor, and accelerator beam neutrinos \cite{SK, SNO, MINOS, KAM, BOR, K2K, DB+RENO}. The underlying mechanism can be described by the unitary PMNS mixing matrix \cite{PMNS:1962}, which connects the three weak flavor eigenstates $(\upnu_{\text{e}},\upnu_{\upmu},\upnu_{\uptau})$ with the three mass eigenstates  $(\upnu_1,\upnu_2,\upnu_3)$. This matrix can be parameterized by three mixing angles, $\theta_{12}$, $\theta_{13}$, and $\theta_{23}$, and a CP-violating phase $\delta_{\text{CP}}$. Oscillation probabilities in vacuum depend upon the mixing parameters, neutrino energy, travel distance (baseline), and the squared neutrino mass differences $\Delta m^2_{ij} \equiv m^2_i - m^2_j$ ($i,j=1,2,3$). Oscillation probabilities in long-baseline experiments can be further modified by matter effects \cite{MSW:1978}.

Current data are well described by the three-flavor model. However, with increasing precision of experiments, one can test for discrepancies that could be accounted for by small modifications to the standard three-flavor model. One such scenario employs large extra dimensions.

Sub-millimeter sized large extra dimensions were originally introduced in \cite{ArkaniHamed:1998rs} to explain the large gap between the electroweak scale, $m_{\text{EW}} \sim 10^3\,\text{GeV}$, and the Planck scale, $M_{\text{Pl}} \sim 10^{19}\,\text{GeV}$. In this model, $M_{\text{Pl}}$ attains its high value due to a volumetric scaling of a more fundamental scale, $\overline{M}_{\text{Pl}}$, which is assumed to be of the same order of magnitude as $m_{\text{EW}}$,
\begin{align}
M_{\text{Pl}}^2 = \overline{M}_{\text{Pl}}^{d+2} V_d, 
  \label{eq:planckscale}
\end{align}
where $d$ is the number of extra dimensions and $V_d$ the corresponding volume. In this framework, however, the lack of a higher fundamental scale disqualifies the see-saw mechanism \cite{see-saw} as an explanation of the small neutrino masses. To resolve this, the existence of sterile neutrinos, arising as Kaluza-Klein (KK) states in the extra dimensions, is suggested in \cite{Dienes,ArkaniHamed:1998vp}, leading to small Dirac neutrino masses \cite{ArkaniHamed:1998vp}
\begin{align}
m_{\upnu} = \kappa v \frac{\overline{M}_{\text{Pl}}}{M_{\text{Pl}}},
  \label{eq:diracmass}
\end{align}
where $\kappa$ is a Yukawa coupling coefficient and $v$ the Higgs vacuum expectation value.

Adopting the Large Extra Dimension (LED) model of \cite{Dvali,Barbieri,Mohapatra,Davoudiasl:2002fq,Machado:2011jt}, all the Standard Model (SM) fields, including the three left-handed (active) neutrinos and the Higgs doublet, live on a four-dimensional brane, 3+1 spacetime. Three SM singlet fermion fields, one for each neutrino flavor, live in a higher-dimensional bulk, 3+1+$d$ spacetime, with at least two compactified extra dimensions ($d \ge 2$). To simplify matters, one of the extra dimensions can be compactified on a circle with radius $R$ much larger than the size of the other dimensions, effectively making this a five-dimensional problem. The compactness of the extra dimension allows a decomposition of each bulk fermion in Fourier modes. 
From the couplings to gauge bosons, the zero modes can be identified as the active neutrinos, while the other modes are sterile neutrinos. All these states are collectively referred to as the KK towers. The Yukawa coupling between the bulk fermions and the active neutrinos leads to mixing between the SM and KK neutrinos, which alters the three-flavor oscillation probabilities. Hence, neutrino oscillation measurements can constrain the size of large extra dimensions. 

As discussed in \cite{Machado:2011jt}, the oscillation amplitude among active neutrino states can be written as
\begin{align}
  \mathcal{A} \left( \upnu_\alpha {\rightarrow} \upnu_\beta \right) = 
  &\sum_{i,j,k = 1}^3 \sum_{n = 0}^{+\infty}
  U_{\alpha i} U_{\beta k}^*    
  W^{(0 n) *}_{ij} W^{(0 n)}_{kj} \nonumber \\
  & \times \exp{\left[ i  \left( \frac{\lambda_j^{(n)}}{R} \right)^2 \left( \frac{L}{2E} \right) \right] },  
  \label{eq:amplitudeLED}
\end{align}
where $E$ is the neutrino energy and $L$ the baseline. The eigenvalues $\lambda^{(n)}_j$ of the Hamiltonian depend on $R$ and the active neutrino masses $m_1$, $m_2$, and $m_3$. The matrices $U$ and $W$ are the mixing matrices for the active and KK neutrino modes, respectively. The $(0n)$ indices refer to the mixing between the zero mode and the KK tower. In practice, only the first five KK modes are considered in each tower \cite{Machado:2011jt}. Squaring the amplitude gives the oscillation probability $P(\upnu_\alpha {\rightarrow} \upnu_\beta)$. Compared to the three-flavor case, this model requires two extra parameters, $R$ and $m_0$, where the latter is defined as the lightest active neutrino mass (normal mass ordering: $m_3 > m_2 > m_1 \equiv m_0$; inverted mass ordering: $m_2 > m_1 > m_3 \equiv m_0$).

\begin{figure}[tbp]
\includegraphics[scale=0.425]{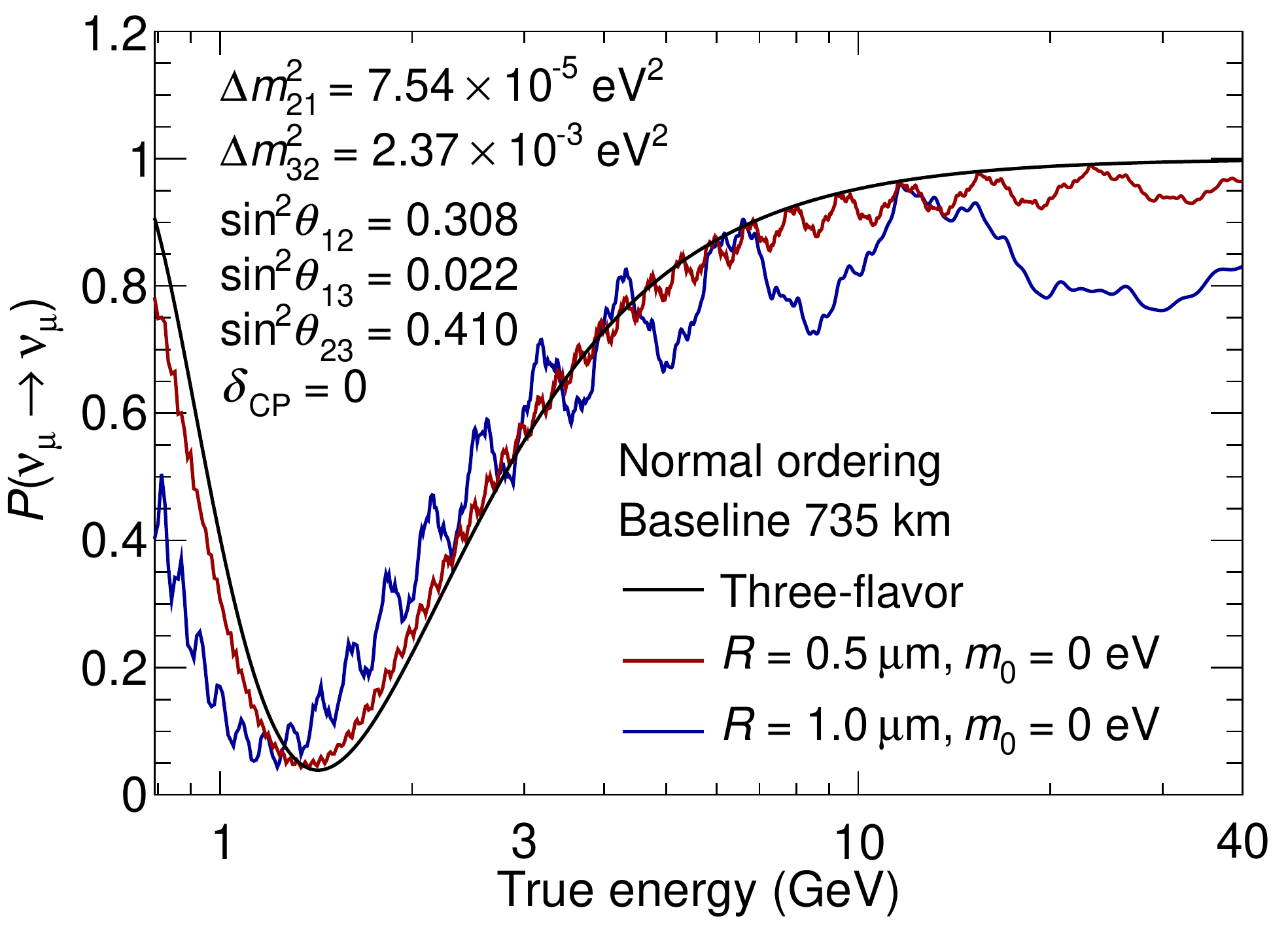}
\caption{The muon neutrino survival probability $P(\upnu_{\upmu} {\rightarrow} \upnu_{\upmu})$ at the MINOS Far Detector as a function of the true neutrino energy for $m_0 = 0\,\text{eV}$ and $R = 0.5\,\upmu\text{m}$ (red line) or $1\,\upmu\text{m}$ (blue line), and for three-flavor oscillation (black line).}
\label{fig:fd_oscillation}
\end{figure}

In Fig.\,\ref{fig:fd_oscillation}, the muon neutrino survival probability, $P(\upnu_{\upmu} {\rightarrow} \upnu_{\upmu})$, for the MINOS baseline and normal mass ordering, is illustrated for $m_0 = 0\,\text{eV}$ and two values of $R$. As stated in \cite{Machado:2011jt}, there are three prominent features of LED visible in this figure: a displacement of the oscillation minimum with respect to the three-flavor case, a reduction of the integrated survival probability because of active-to-KK oscillation, and the appearance of modulations on the survival probability because of fast oscillations to the KK states. With increasing energy, the amplitude of the modulations increases while their frequency decreases, making the effects of LED easier to observe away from the oscillation minimum. The values in Fig.\,\ref{fig:fd_oscillation} of $\Delta m^2_{32} = 2.37 \times 10^{-3}\,\text{eV}^2$ and $\sin^2 \theta_{32} = 0.410$ are taken from the MINOS standard oscillation analysis \cite{MINOS}. The value of $\sin^2 \theta_{13} = 0.022$ is a weighted average of the Daya Bay \cite{DayaBayPars}, RENO \cite{RENOPars}, and Double Chooz \cite{DoubleChoozPars} results. The values of $\Delta m^2_{21} = 7.54 \times 10^{-5}\,\text{eV}^2$ and $\sin^2 \theta_{12} = 0.308$ are taken from a global fit \cite{GlobalFitPars}. We set $\delta_{\text{CP}} = 0$ since it has little effect on the oscillation probabilities \cite{Junting}. These values are used throughout the analysis.

\begin{figure}[tbp]
\includegraphics[scale=0.425]{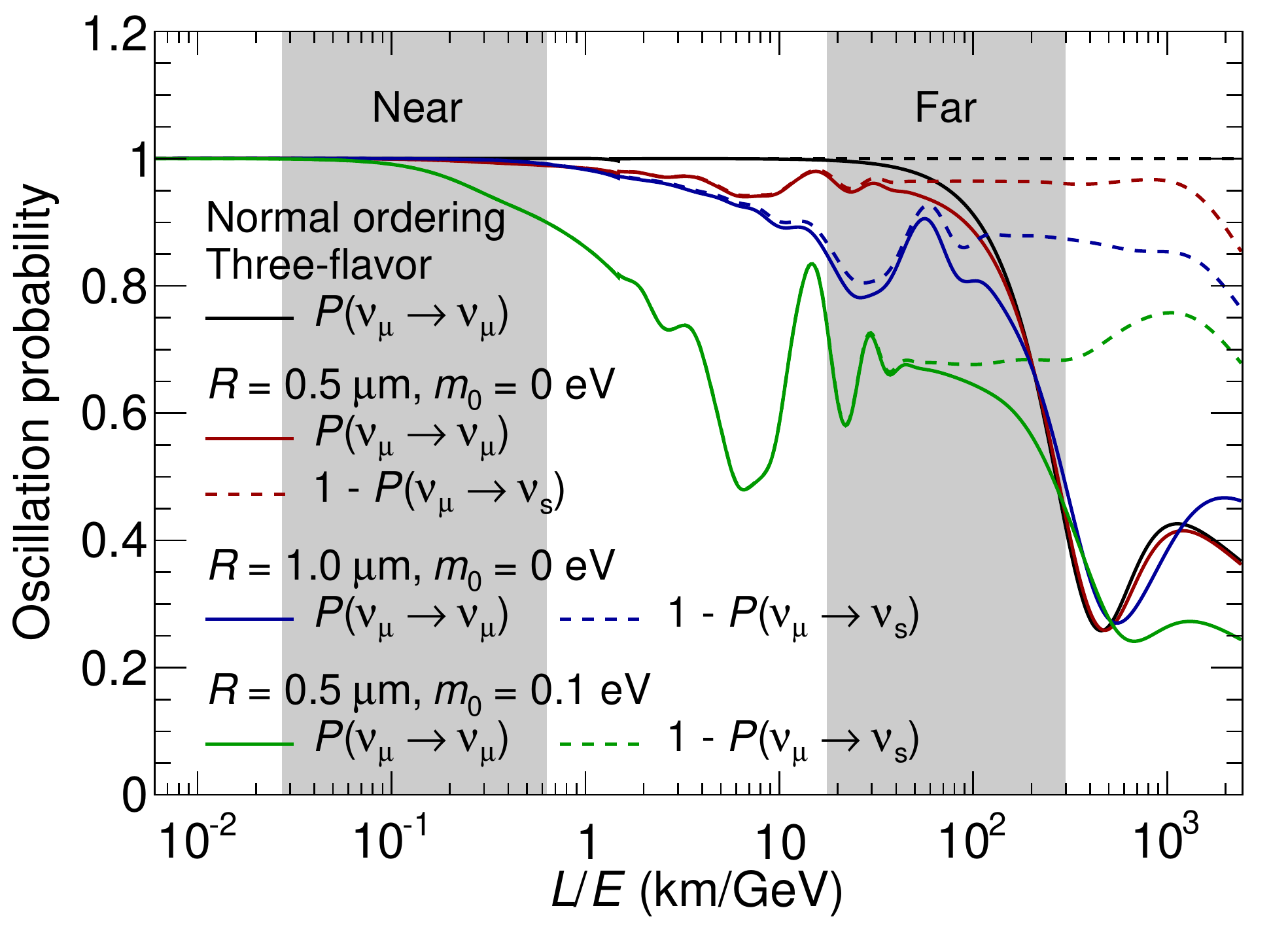}
\caption{The oscillation probabilities as a function of $L/E$, incorporating MINOS energy resolution effects, for the three-flavor case (black line), the same LED scenarios as in Fig.\,\ref{fig:fd_oscillation} (red and blue lines), and an LED case with non-zero $m_0$ (green lines). The baseline and neutrino energy are denoted as $L$ and $E$, respectively. The $L/E$ coverage of the Near Detector and Far Detector are represented by the grey bands, which contain $90\%$ of the MINOS data.}
\label{fig:l_over_e}
\end{figure}

Constraints on this model based on atmospheric, reactor, and accelerator beam neutrino oscillation experiments are discussed in \cite{Davoudiasl:2002fq}. The authors derived a bound constraining $R$ to be less than $0.82\,\upmu\text{m}$ at 90\%\,C.L. The sensitivities of CHOOZ, KamLAND, and MINOS  for this model were calculated in \cite{Machado:2011jt} using a modified version of the GLoBES software \cite{GLoBES}. Assuming $7.24 \times 10^{20}$ protons-on-target (POT) for MINOS in $\upnu_{\upmu}$ mode, a combined sensitivity of $R < 0.75\,(0.49)\,\upmu\text{m}$ at 90\%\,C.L. was obtained for normal (inverted) mass ordering and vanishing $m_0$. The zenith distribution of atmospheric neutrino events collected by IceCube was analyzed in \cite{Esmaili1}, where the authors showed that it is possible to exclude $R \gtrsim 0.40\,\upmu\text{m}$ at 95\%\,C.L. The above neutrino limits are about two orders of magnitude stronger than those obtained from tabletop gravitational experiments \cite{PDG:2014}. Astrophysical and cosmological bounds are often much more stringent, but are model-dependent \cite{PDG:2014}. Collider experiments can set limits on the volume of the extra dimensions \cite{PDG:2014}. This paper presents an analysis, sensitive to LED, of the MINOS data set with a low-energy $\upnu_{\upmu}$ mode exposure of $10.56\times10^{20}$ POT and a peak neutrino energy of $3\,\text{GeV}$.

In the MINOS experiment, neutrinos are produced by directing $120\,\text{GeV}$ protons from the Fermilab Main Injector onto a graphite target. The resulting $\uppi$ and K mesons are focused in the forward direction and charge-sign selected by two magnetic horns after which they decay into neutrinos in a $675\,\text{m}$ long tunnel. MINOS observes charged current (CC) and neutral current (NC) neutrino events in a Near Detector (ND) and Far Detector (FD) located $1.04\,\text{km}$ and $735\,\text{km}$ downstream of the target, respectively, on the NuMI (Neutrinos at the Main Injector) beamline axis \cite{Adamson:2016beam}. The beam composition at the ND consists of $91.8\%$ $\upnu_{\upmu}$, $6.9\%$ $\overline{\upnu}_{\upmu}$, and $1.3\%$ $\upnu_{\text{e}}$ and $\overline{\upnu}_{\text{e}}$ when operating in $\upnu_{\upmu}$ mode. The ND and FD are tracking-sampling calorimeters built of $2.54\,\text{cm}$ thick iron plates interleaved with scintillator planes composed of $1\,\text{cm}$ thick, $4.1\,\text{cm}$ wide strips, arranged in two alternating orthogonal views, and read out using wavelength-shifting fibers coupled to multi-anode photomultiplier tubes. The ND has a $23.7\,\text{t}$ fiducial ($980\,\text{t}$ total) mass. The FD has a $4.2\,\text{kt}$ fiducial ($5.4\,\text{kt}$ total) mass. Using magnetic coils, both detectors are magnetized with a toroidal magnetic field oriented to focus negatively charged particles when operating in $\upnu_{\upmu}$ mode \cite{Michael:2008bc}.

A $\upnu_{\upmu}$ CC-like event ($\upnu_{\upmu} \text{N} {\rightarrow} \upmu \text{X}$) in the MINOS detectors is characterized by a single outgoing muon track with possible hadronic showers near the event vertex. The muon momentum is determined from the track range for tracks confined within the detector and from the track curvature for exiting tracks. Since no charge separation is applied in this analysis, both $\upnu_{\upmu}$ and $\overline{\upnu}_{\upmu}$ events are used. The energy of CC hadronic showers is estimated using a $k$-nearest-neighbor ($k$NN) algorithm based on the shower topology, in addition to the calorimetric shower energy \cite{kNN:1,kNN:2,kNN:3,kNN:4,kNN:5}. The CC neutrino energy at the FD is reconstructed with a mean resolution of $17.3\%$ by summing the track and shower energies.
 
A $\upnu_{\upalpha}$ NC-like event ($\upnu_{\upalpha} \text{N} {\rightarrow} \upnu_{\upalpha} \text{X}\text{ with }\alpha=\text{e},\upmu,\uptau$) in the ND or FD has a short diffuse hadronic shower and possibly short hadron tracks. The NC event length is required to be shorter than 47 planes. If a hadron track is reconstructed in an event, the track length is required not to exceed the shower length by more than five planes. Additional selection requirements are imposed in the ND to remove cases where reconstruction failed due to high event rates \cite{Gemma,sterile2011}. The NC neutrino energy at the FD is reconstructed with a mean resolution of $41.7\%$ using the calorimetric shower energy.

In the CC selection procedure, separation between CC and NC events is performed by combining four variables describing track properties into a single discriminant variable using a $k$NN algorithm \cite{kNN:6,kNN:7}. Only events that failed the NC selection procedure are considered in the CC selection procedure. The selected CC sample has an efficiency (purity) estimated by Monte Carlo (MC) of $53.9\%$ ($98.7\%$) at the ND and $84.6\%$ ($99.1\%$) at the FD. The ND efficiency is low because events occurring near the magnetic coil hole are rejected. NC events are the main background in both detectors \cite{Junting}.

The selected NC sample has an efficiency (purity) estimated by MC of $79.9\%$ ($58.9\%$) at the ND and $87.6\%$ ($61.3\%$) at the FD. The background composition is $86.9\%$ CC $\upnu_{\upmu}$ and $13.1\%$ beam CC $\upnu_{\text{e}}$ at the ND. Assuming three-flavor oscillation, the backgrounds at the FD are estimated as $73.8\%$ CC $\upnu_{\upmu}$, $21.6\%$ CC $\upnu_{\text{e}}$, and $4.6\%$ CC $\upnu_{\uptau}$ \cite{Junting}. 

The muon neutrino survival probability $P(\upnu_{\upmu} {\rightarrow} \upnu_{\upmu})$, probed by CC events, and the sterile neutrino appearance probability $P(\upnu_{\upmu} {\rightarrow} \upnu_{\text{s}})$, probed by a depletion of NC events, are shown as a function of $L/E$ in Fig.\,\ref{fig:l_over_e}. The same two LED scenarios as in Fig.\,\ref{fig:fd_oscillation} and a scenario with non-zero $m_0$ are compared to the three-flavor case. MINOS energy resolution effects \cite{Kordosky,smearing} were accounted for when calculating the probabilities. 

\begin{figure}[tp]
\includegraphics[scale=0.425]{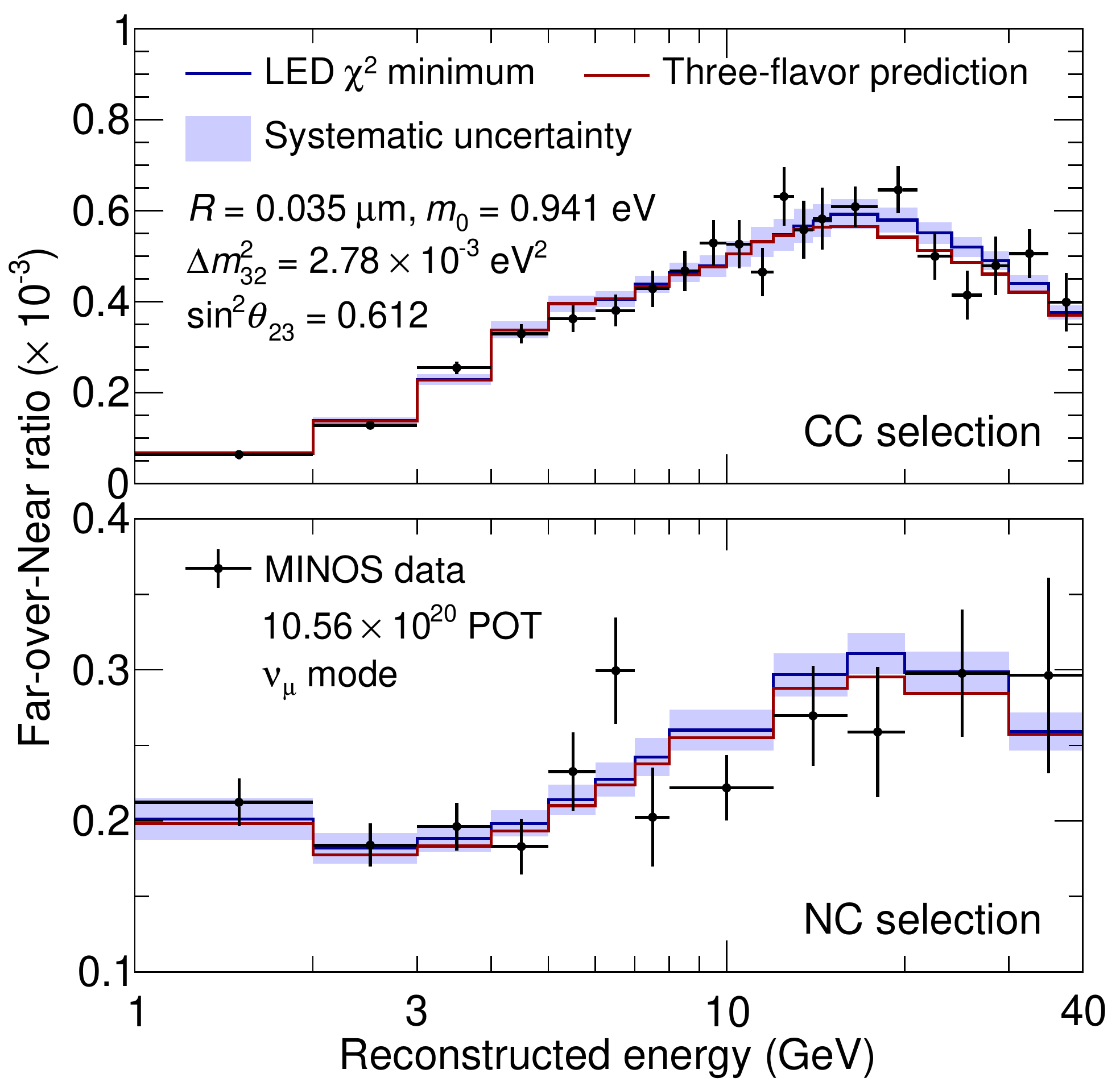}
\caption{The Far-over-Near ratio of the MINOS CC (top) and NC (bottom) data (black points) as a function of the reconstructed neutrino energy. The error bars represent statistical uncertainties. The three-flavor predictions, using the same oscillation parameters as Fig.\,\ref{fig:fd_oscillation}, are shown in red. The LED predictions of the $\chi^2$ minimum are shown with their systematic uncertainty (blue line and band).}
\label{fig:data}
\end{figure}

In the MINOS standard oscillation analysis \cite{MINOS}, ND data are used to constrain the FD prediction based on the assumption of no oscillations along the ND baseline (black line in Fig.\,\ref{fig:l_over_e}). In the LED analysis, this assumption is not valid for $m_0 \gtrsim 30\,\text{meV}$ when $R \gtrsim 0.01\,\upmu\text{m}$ (as illustrated by the green lines in Fig.\,\ref{fig:l_over_e}), and a fit is performed to the ratio of the reconstructed FD and ND neutrino energy spectra. This Far-over-Near ratio fit, using a covariance-matrix-based $\chi^2$ method \cite{Junting}, allows the analysis to be sensitive to oscillations along the ND and FD baselines and significantly reduces many systematics affecting both detectors. Figure \ref{fig:data} compares the Far-over-Near ratio of the MINOS CC and NC data to the three-flavor predictions and the LED predictions of the $\chi^2$ minimum. The good agreement between data and the three-flavor predictions indicates that, if large extra dimensions exist, oscillation between active and KK states must be sub-dominant in MINOS. 

The energy window for the fit is set between $0$ and $40\,\text{GeV}$, with the CC and NC binning schemes chosen such that the minimum number of FD events in a bin provides a good Gaussian approximation. The CC and NC samples are fitted simultaneously to improve the sensitivity. The total $\chi^2$ is the sum of those of the CC and NC samples, with each one given by
\begin{align}
  \chi^2 = \sum_{i,j = 1}^N (o_i - p_i) [V^{-1}]_{ij} (o_j - p_j) 
  + \left(\frac{N_{\text{data}} - N_{\text{MC}}}{\sigma_N}\right)^2,
  \label{eq:chisqForLED}
\end{align}
where $o_i$ and $p_i$ are the observed and predicted Far-over-Near ratios in energy bin $i$, respectively, and $V$ is the sum of statistical and systematic covariance matrices. The second term is an ND beam flux penalty term, where $N_{\text{data}}$ ($N_{\text{MC}}$) is the total number of ND data (MC) events, and $\sigma_N = 50\%\,N_{\text{MC}}$ is adopted as a conservative difference between hadron production measurements and MC calculations \cite{Junting}.

The total covariance matrix is
\begin{align}
V = & \ V_{\text{stat}} + V_{\text{acc}} + V_{\text{norm}} + V_{\text{NC}} + V_{\text{other}}.
\label{eq:covariance}
\end{align}
The statistical uncertainties are contained in $V_{\text{stat}}$ and are less than $24\%$ in each energy bin and $15\%$ on average. The acceptance ($V_{\text{acc}}$), normalization ($V_{\text{norm}}$), and NC selection  ($V_{\text{NC}}$) covariance matrices have the biggest impact on the sensitivity and are discussed below. Other systematic uncertainties ($V_{\text{other}}$), including neutrino interaction cross-section uncertainties and NuMI beam flux uncertainties, are small and have a cumulative effect of less than $4\%$ in any energy bin of the Far-over-Near ratio. 

The uncertainty in the acceptance and efficiency of the ND for CC and NC events is evaluated by varying event selection requirements in data and MC. Any shift in the data-MC agreement is taken as a systematic uncertainty. The effect on the Far-over-Near ratio of this systematic uncertainty is energy-dependent, never exceeding $6\%$ ($0.6\%$) for the CC (NC) sample, and includes correlations between different bins.

The normalization systematic uncertainty is a consequence of the detector differences between ND and FD, including material dimensions, detector live time, and reconstruction efficiencies. It has a uniform uncertainty in the Far-over-Near ratio of $1.6\%$ $(2.2\%)$ for the CC (NC) sample. 

The matrix $V_{\text{NC}}$ accounts for the uncertainty in the selection procedure that reduces the number of poorly-reconstructed NC events, defined as those events with reconstructed energy less than $30\%$ of the true energy. To improve data-MC agreement, the fraction of poorly reconstructed events in the simulation is varied in a template fit to the selection variables. The selection criteria are then adjusted to yield the same number of rejected events in data and MC. The variations seen in the NC energy spectra from this procedure are taken as a systematic uncertainty. In the Far-over-Near ratio, this uncertainty varies from $5\%$ at $1\,\text{GeV}$ to $1\%$ at $10\,\text{GeV}$.

\begin{figure}[tbp]
\includegraphics[scale=0.425]{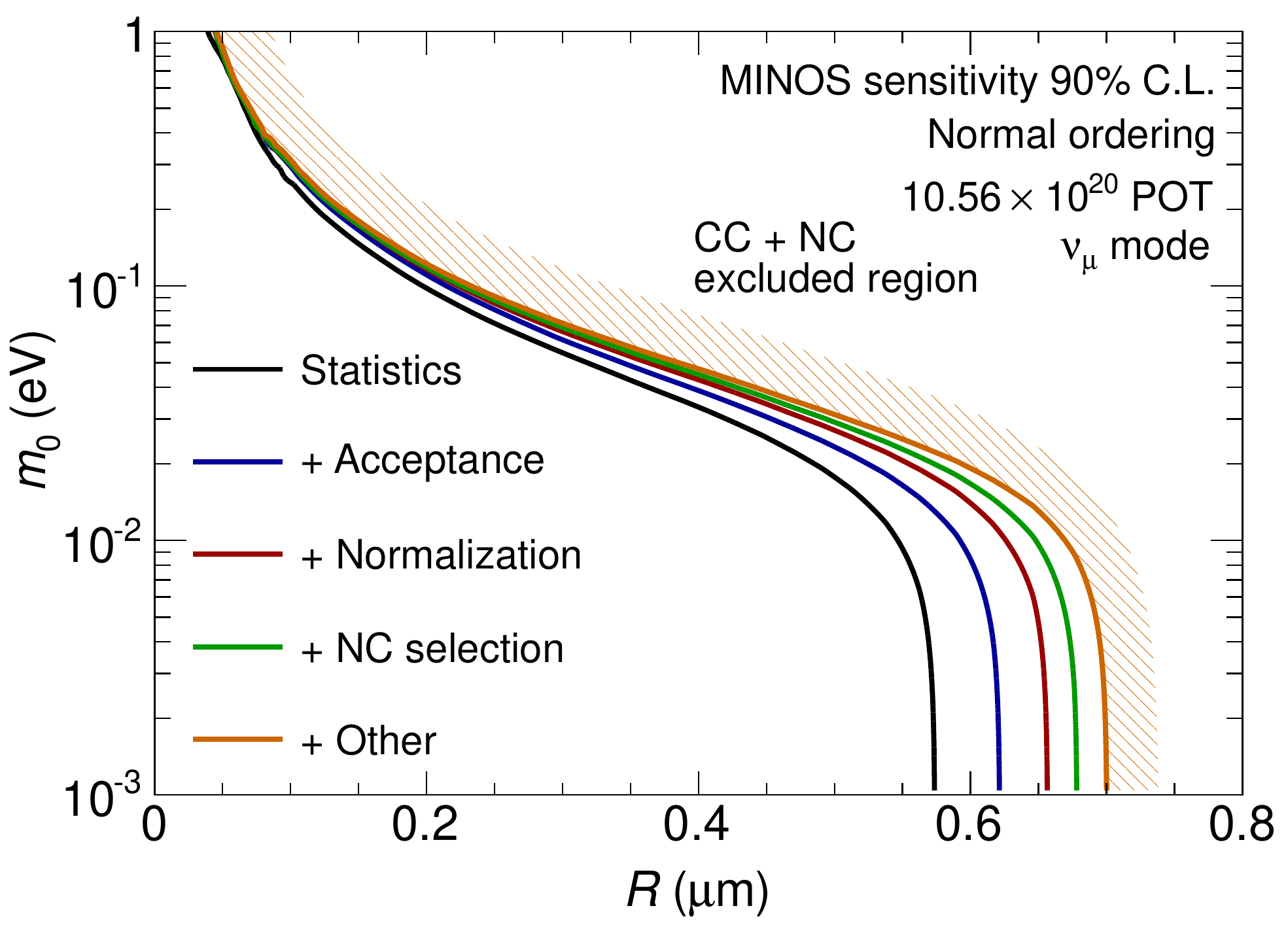}
\caption{The cumulative effect of the systematic uncertainties on the $90\%$\,C.L. sensitivity contour based on $10.56\times10^{20}$ POT MC and assuming normal mass ordering. The large extra dimension size and the smallest neutrino mass are denoted as $R$ and $m_0$, respectively. The shaded area indicates the excluded region to the right of the contour.}
\label{fig:systematics}
\end{figure}

In minimizing the $\chi^2$ in the $(R,m_0)$ plane, $\theta_{23}$ and $\Delta m^2_{32}$ are free to vary in the fit. The four-dimensional parameter space is divided into $51 {\times} 51 {\times} 26 {\times} 51$ bins and has ranges $\left[10^{-8},10^{-6}\right]\text{m}$, $\left[10^{-3},1\right]\text{eV}$, $\left[0,\pi/2\right]$, and $\left[0,5\times10^{-3}\right]\text{eV}^2$ for $R$, $m_0$, $\theta_{23}$, and $\Delta m^2_{32}$, respectively. The Far-over-Near ratio is calculated at each bin center and multilinear interpolation is used to obtain the Far-over-Near ratio at other points in the parameter space. Two initial $\theta_{23}$ hypotheses, one in each octant, are used in the fit. Since the mass ordering was shown to have little effect on the MINOS sensitivity \cite{Machado:2011jt,Junting}, only normal ordering is considered in this analysis. The parameters $\Delta m^2_{21}$, $\theta_{12}$, $\theta_{13}$, and $\delta_{\text{CP}}$ are fixed to the values shown in Fig.\,\ref{fig:fd_oscillation}. CPT symmetry is assumed, implying identical $\upnu$ and $\overline{\upnu}$ oscillation parameters \cite{CPT:1,CPT:2}. The $90\%$\,C.L. sensitivity contour and the cumulative effect of the systematic uncertainties are shown in Fig.\,\ref{fig:systematics}. The sensitivity is calculated using simulated three-flavor data generated for the oscillation parameter values shown in Fig.\,\ref{fig:fd_oscillation}.

\begin{figure}[tbp]
\includegraphics[scale=0.425]{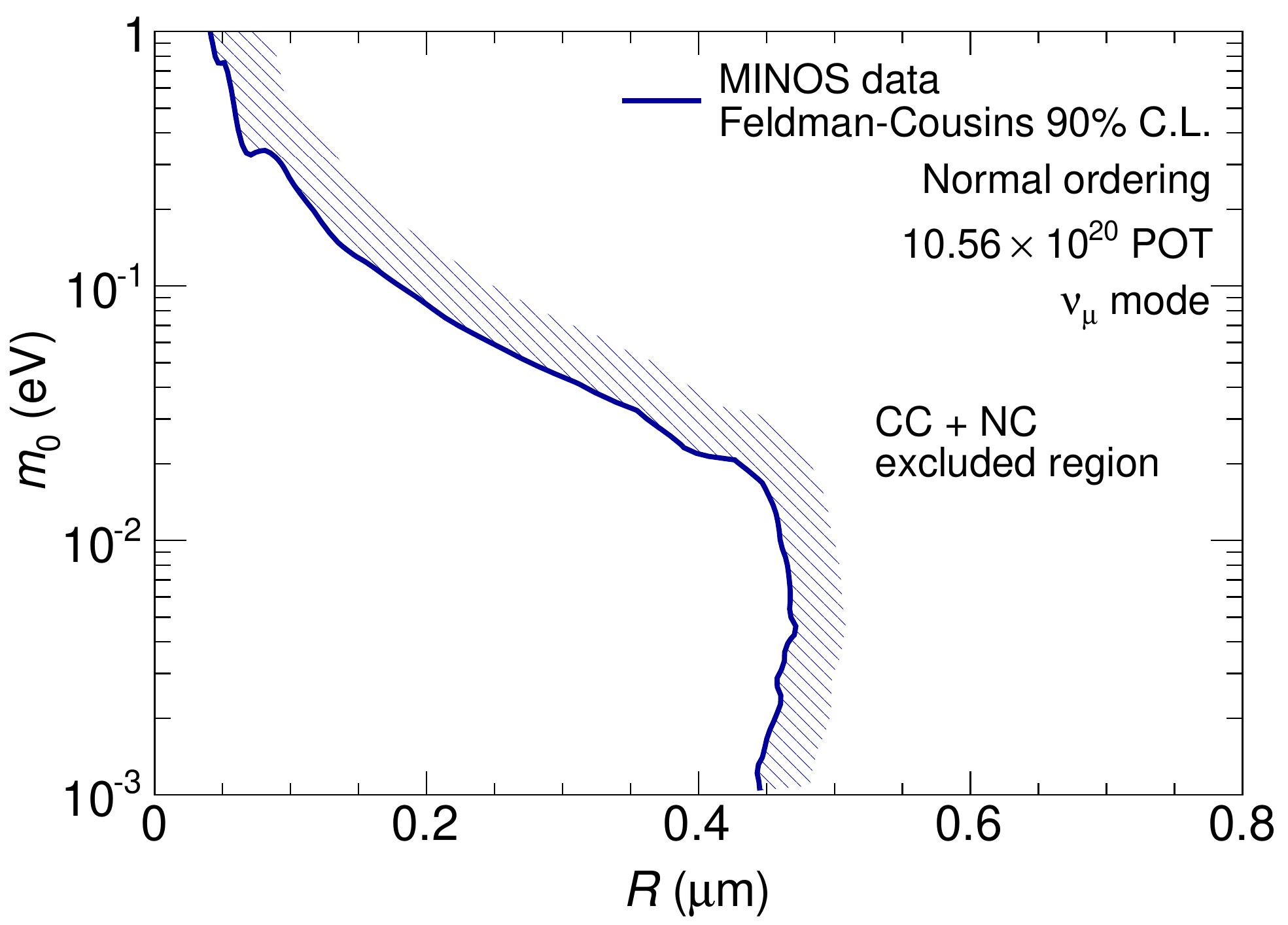}
\caption{The $90\%$\,C.L. data contour for the LED model, obtained using the Feldman-Cousins technique, based on $10.56\times10^{20}$ POT MINOS data and assuming normal mass ordering. The large extra dimension size and the smallest neutrino mass are denoted as $R$ and $m_0$, respectively. The shaded area indicates the excluded region to the right of the contour.}
\label{fig:result}
\end{figure}

The Feldman-Cousins technique \cite{FC:1998} is used to obtain the $90\%$\,C.L. data contour shown in Fig.\,\ref{fig:result}. A shallow global minimum is found at $R = 0.035\,\text{$\upmu$m}$, $m_0 = 0.941\,\text{eV}$, $\sin^2 \theta_{23} = 0.612$, and $\Delta m^2_{32} = 2.78 \times 10^{-3}\,\text{eV}^{2}$ with $\chi^2_{\text{3-flavor}} - \chi^2_{\text{LED}} = 0.95$. No evidence for large extra dimensions is found. The limit obtained from the data is stronger than expected from the sensitivity, as can be seen from a comparison of Figs.\,\ref{fig:systematics} and \ref{fig:result}. A study of 1000 simulated experiments, each one using a Gaussian fluctuation of the simulated three-flavor data based on the full covariance matrices, shows that 39\% of simulated experiments obtain an exclusion stronger than that obtained from the data at $m_0 = 0.005\,\text{eV}$. In the limit of a vanishing lightest neutrino mass, the large extra dimension size is constrained to be smaller than $0.45\,\upmu\text{m}$ at $90\%$\,C.L. To date, this is the strongest limit on this large extra dimension model \cite{Davoudiasl:2002fq,Machado:2011jt} reported by a neutrino oscillation experiment.

\input led_acknowledgement.tex

\end{document}

%% file: minos-plus_feb2016.tex
\newcommand{\Berkeley}{Lawrence Berkeley National Laboratory, Berkeley, California, 94720 USA}
\newcommand{\Cambridge}{Cavendish Laboratory, University of Cambridge, 
Cambridge CB3 0HE, United Kingdom}
\newcommand{\Cincinnati}{Department of Physics, University of Cincinnati, Cincinnati, Ohio 45221, USA}
\newcommand{\FNAL}{Fermi National Accelerator Laboratory, Batavia, Illinois 60510, USA}
\newcommand{\RAL}{Rutherford Appleton Laboratory, Science and Technology Facilities Council, Didcot, OX11 0QX, United Kingdom}
\newcommand{\UCL}{Department of Physics and Astronomy, University College London, 
London WC1E 6BT, United Kingdom}
\newcommand{\Caltech}{Lauritsen Laboratory, California Institute of Technology, Pasadena, California 91125, USA}
\newcommand{\Alabama}{Department of Physics and Astronomy, University of Alabama, Tuscaloosa, Alabama 35487, USA}
\newcommand{\ANL}{Argonne National Laboratory, Argonne, Illinois 60439, USA}
\newcommand{\Athens}{Department of Physics, University of Athens, GR-15771 Athens, Greece}
\newcommand{\NTUAthens}{Department of Physics, National Tech. University of Athens, GR-15780 Athens, Greece}
\newcommand{\Benedictine}{Physics Department, Benedictine University, Lisle, Illinois 60532, USA}
\newcommand{\BNL}{Brookhaven National Laboratory, Upton, New York 11973, USA}
\newcommand{\CdF}{APC -- Universit\'{e} Paris 7 Denis Diderot, 10, rue Alice Domon et L\'{e}onie Duquet, F-75205 Paris Cedex 13, France}
\newcommand{\Cleveland}{Cleveland Clinic, Cleveland, Ohio 44195, USA}
\newcommand{\Delhi}{Department of Physics \& Astrophysics, University of Delhi, Delhi 110007, India}
\newcommand{\GEHealth}{GE Healthcare, Florence South Carolina 29501, USA}
\newcommand{\Harvard}{Department of Physics, Harvard University, Cambridge, Massachusetts 02138, USA}
\newcommand{\HolyCross}{Holy Cross College, Notre Dame, Indiana 46556, USA}
\newcommand{\Houston}{Department of Physics, University of Houston, Houston, Texas 77204, USA}
\newcommand{\IIT}{Department of Physics, Illinois Institute of Technology, Chicago, Illinois 60616, USA}
\newcommand{\Iowa}{Department of Physics and Astronomy, Iowa State University, Ames, Iowa 50011 USA}
\newcommand{\Indiana}{Indiana University, Bloomington, Indiana 47405, USA}
\newcommand{\ITEP}{High Energy Experimental Physics Department, ITEP, B. Cheremushkinskaya, 25, 117218 Moscow, Russia}
\newcommand{\JMU}{Physics Department, James Madison University, Harrisonburg, Virginia 22807, USA}
\newcommand{\LASL}{Nuclear Nonproliferation Division, Threat Reduction Directorate, Los Alamos National Laboratory, Los Alamos, New Mexico 87545, USA}
\newcommand{\Lebedev}{Nuclear Physics Department, Lebedev Physical Institute, Leninsky Prospect 53, 119991 Moscow, Russia}
\newcommand{\Lancaster}{Lancaster University, Lancaster, LA1 4YB, UK}
\newcommand{\LLL}{Lawrence Livermore National Laboratory, Livermore, California 94550, USA}
\newcommand{\LosAlamos}{Los Alamos National Laboratory, Los Alamos, New Mexico 87545, USA}
\newcommand{\Manchester}{School of Physics and Astronomy, University of Manchester, 
Manchester M13 9PL, United Kingdom}
\newcommand{\MIT}{Lincoln Laboratory, Massachusetts Institute of Technology, Lexington, Massachusetts 02420, USA}
\newcommand{\Minnesota}{University of Minnesota, Minneapolis, Minnesota 55455, USA}
\newcommand{\Crookston}{Math, Science and Technology Department, University of Minnesota -- Crookston, Crookston, Minnesota 56716, USA}
\newcommand{\Duluth}{Department of Physics, University of Minnesota Duluth, Duluth, Minnesota 55812, USA}
\newcommand{\Ohio}{Center for Cosmology and Astro Particle Physics, Ohio State University, Columbus, Ohio 43210 USA}
\newcommand{\Otterbein}{Otterbein University, Westerville, Ohio 43081, USA}
\newcommand{\Oxford}{Subdepartment of Particle Physics, University of Oxford, Oxford OX1 3RH, United Kingdom}
\newcommand{\PennState}{Department of Physics, Pennsylvania State University, State College, Pennsylvania 16802, USA}
\newcommand{\PennU}{Department of Physics and Astronomy, University of Pennsylvania, Philadelphia, Pennsylvania 19104, USA}
\newcommand{\Pittsburgh}{Department of Physics and Astronomy, University of Pittsburgh, Pittsburgh, Pennsylvania 15260, USA}
\newcommand{\IHEP}{Institute for High Energy Physics, Protvino, Moscow Region RU-140284, Russia}
\newcommand{\Rochester}{Department of Physics and Astronomy, University of Rochester, New York 14627 USA}
\newcommand{\RoyalH}{Physics Department, Royal Holloway, University of London, Egham, Surrey, TW20 0EX, United Kingdom}
\newcommand{\Carolina}{Department of Physics and Astronomy, University of South Carolina, Columbia, South Carolina 29208, USA}
\newcommand{\SDakota}{South Dakota School of Mines and Technology, Rapid City, South Dakota 57701, USA}
\newcommand{\SLAC}{Stanford Linear Accelerator Center, Stanford, California 94309, USA}
\newcommand{\Stanford}{Department of Physics, Stanford University, Stanford, California 94305, USA}
\newcommand{\StJohnFisher}{Physics Department, St. John Fisher College, Rochester, New York 14618 USA}
\newcommand{\Sussex}{Department of Physics and Astronomy, University of Sussex, Falmer, Brighton BN1 9QH, United Kingdom}
\newcommand{\TexasAM}{Physics Department, Texas A\&M University, College Station, Texas 77843, USA}
\newcommand{\Texas}{Department of Physics, University of Texas at Austin, 
Austin, Texas 78712, USA}
\newcommand{\TechX}{Tech-X Corporation, Boulder, Colorado 80303, USA}
\newcommand{\Tufts}{Physics Department, Tufts University, Medford, Massachusetts 02155, USA}
\newcommand{\UNICAMP}{Universidade Estadual de Campinas, IFGW, CP 6165, 13083-970, Campinas, SP, Brazil}
\newcommand{\UFG}{Instituto de F\'{i}sica, 
Universidade Federal de Goi\'{a}s, 74690-900, Goi\^{a}nia, GO, Brazil}
\newcommand{\USP}{Instituto de F\'{i}sica, Universidade de S\~{a}o Paulo,  CP 66318, 05315-970, S\~{a}o Paulo, SP, Brazil}
\newcommand{\Warsaw}{Department of Physics, University of Warsaw, 
PL-02-093 Warsaw, Poland}
\newcommand{\Washington}{Physics Department, Western Washington University, Bellingham, Washington 98225, USA}
\newcommand{\WandM}{Department of Physics, College of William \& Mary, Williamsburg, Virginia 23187, USA}
\newcommand{\Wisconsin}{Physics Department, University of Wisconsin, Madison, Wisconsin 53706, USA}
\newcommand{\deceased}{Deceased.}

\affiliation{\ANL}
\affiliation{\Athens}
\affiliation{\BNL}
\affiliation{\Caltech}
\affiliation{\Cambridge}
\affiliation{\UNICAMP}
\affiliation{\Cincinnati}
\affiliation{\FNAL}
\affiliation{\UFG}
\affiliation{\Harvard}
\affiliation{\HolyCross}
\affiliation{\Houston}
\affiliation{\IIT}
\affiliation{\Indiana}
\affiliation{\Iowa}
\affiliation{\Lancaster}
\affiliation{\UCL}
\affiliation{\Manchester}
\affiliation{\Minnesota}
\affiliation{\Duluth}
\affiliation{\Otterbein}
\affiliation{\Oxford}
\affiliation{\Pittsburgh}
\affiliation{\RAL}
\affiliation{\USP}
\affiliation{\Carolina}
\affiliation{\Stanford}
\affiliation{\Sussex}
\affiliation{\TexasAM}
\affiliation{\Texas}
\affiliation{\Tufts}
\affiliation{\Warsaw}
\affiliation{\WandM}

\author{P.~Adamson}
\affiliation{\FNAL}


\author{I.~Anghel}
\affiliation{\Iowa}
\affiliation{\ANL}



\author{A.~Aurisano}
\affiliation{\Cincinnati}









\author{G.~Barr}
\affiliation{\Oxford}









\author{M.~Bishai}
\affiliation{\BNL}

\author{A.~Blake}
\affiliation{\Cambridge}
\affiliation{\Lancaster}


\author{G.~J.~Bock}
\affiliation{\FNAL}


\author{D.~Bogert}
\affiliation{\FNAL}




\author{S.~V.~Cao}
\affiliation{\Texas}

\author{T.~J.~Carroll}
\affiliation{\Texas}

% ...

\collaboration{The MINOS Collaboration}
\noaffiliation

%% file: led_acknowledgement.tex
This work was supported by the U.S. DOE; the United Kingdom STFC; the U.S. NSF; the State and University of Minnesota; and Brazil's FAPESP, CNPq and CAPES. We are grateful to the Minnesota Department of Natural Resources and the personnel of the Soudan Laboratory and Fermilab. We thank the Texas Advanced Computing Center at The University of Texas at Austin for the provision of computing resources. We wish to thank P.~A.~N.~Machado for providing insightful comments on the LED model.